\documentclass[longauth]{aa} 

\usepackage{graphicx,latexsym,amssymb}
\usepackage{lscape}
\usepackage{natbib}
\bibpunct{(}{)}{;}{a}{}{,} 
\usepackage{times}

\begin{document}
\newcommand {\xmm} {XMM-{\it Newton}}
\newcommand {\sax} {{\it Beppo}SAX }
\newcommand {\rosat} {{ROSAT }}
\newcommand {\swift} {{\it Swift }}
\newcommand {\hess} {{HESS }}
\newcommand {\rchisq} {$\chi_{\nu} ^{2}$}
\newcommand {\chisq} {$\chi^{2}$}
\newcommand {\ergs}[1]{$\times10^{#1}$ ergs cm$^{-2}$ s$^{-1}$}
\newcommand {\e}[1]{$\;\times10^{#1}$}
\newcommand {\nufnu}{$\nu F_{\nu}$ }
\newcommand {\chandra}{{\it Chandra} }
\newcommand {\nw}{nW m$^{-2}$sr$^{-1}$}
\newcommand {\micron}{$\mu$m }
\newcommand {\cms}{cm$^{-2}$s$^{-1}$ }

\title{PKS\,2005$-$489 at VHE: Four years of monitoring with HESS
   and simultaneous multi-wavelength observations}
\author{HESS Collaboration
 \and F.~Acero \inst{15}
 \and F. Aharonian\inst{1,13}
 \and A.G.~Akhperjanian \inst{2}
 \and G.~Anton \inst{16}
 \and U.~Barres de Almeida \inst{8} \thanks{supported by CAPES Foundation, Ministry of Education of Brazil}
 \and A.R.~Bazer-Bachi \inst{3}
 \and Y.~Becherini \inst{12}
 \and B.~Behera \inst{14}
 \and W.~Benbow \inst{1} \thanks{now at Harvard-Smithsonian Center for Astrophysics, Cambridge, USA}
 \and K.~Bernl\"ohr \inst{1,5}
% ...
}
\offprints{luigi.costamante@stanford.edu or wbenbow@cfa.harvard.edu}
\institute{
Max-Planck-Institut f\"ur Kernphysik, P.O. Box 103980, D 69029
Heidelberg, Germany
\and
 Yerevan Physics Institute, 2 Alikhanian Brothers St., 375036 Yerevan,
Armenia
\and
Centre d'Etude Spatiale des Rayonnements, CNRS/UPS, 9 av. du Colonel Roche, BP
4346, F-31029 Toulouse Cedex 4, France
\and
Universit\"at Hamburg, Institut f\"ur Experimentalphysik, Luruper Chaussee
149, D 22761 Hamburg, Germany
\and
Institut f\"ur Physik, Humboldt-Universit\"at zu Berlin, Newtonstr. 15,
D 12489 Berlin, Germany
\and
LUTH, Observatoire de Paris, CNRS, Universit\'e Paris Diderot, 5 Place Jules Janssen, 92190 Meudon, 
France
\and
IRFU/DSM/CEA, CE Saclay, F-91191
Gif-sur-Yvette, Cedex, France
\and
University of Durham, Department of Physics, South Road, Durham DH1 3LE,
U.K.
\and
Unit for Space Physics, North-West University, Potchefstroom 2520,
    South Africa
\and
Laboratoire Leprince-Ringuet, Ecole Polytechnique, CNRS/IN2P3,
 F-91128 Palaiseau, France
\and 
Laboratoire d'Annecy-le-Vieux de Physique des Particules,
Universit\'{e} de Savoie, CNRS/IN2P3, F-74941 Annecy-le-Vieux,
France
\and
Astroparticule et Cosmologie (APC), CNRS, Universite Paris 7 Denis Diderot,
10, rue Alice Domon et Leonie Duquet, F-75205 Paris Cedex 13, France
\thanks{UMR 7164 (CNRS, Universit\'e Paris VII, CEA, Observatoire de Paris)}
\and
Dublin Institute for Advanced Studies, 5 Merrion Square, Dublin 2,
Ireland
\and
Landessternwarte, Universit\"at Heidelberg, K\"onigstuhl, D 69117 Heidelberg, Germany
\and
Laboratoire de Physique Th\'eorique et Astroparticules, 
Universit\'e Montpellier 2, CNRS/IN2P3, CC 70, Place Eug\`ene Bataillon, F-34095
Montpellier Cedex 5, France
\and
Universit\"at Erlangen-N\"urnberg, Physikalisches Institut, Erwin-Rommel-Str. 1,
D 91058 Erlangen, Germany
% ...
}
   \date{Received; accepted}
%
%
  \abstract
   {}
   {Our aim is to study the very high energy (VHE; $E>100$ GeV) $\gamma$-ray emission
from BL\,Lac objects and the evolution in time of their 
broad-band spectral energy distribution (SED).}
   {VHE observations of the high-frequency peaked BL\,Lac object PKS\,2005$-$489 were made with
the High Energy Stereoscopic System (HESS) from 2004 through 2007.  
Three simultaneous multi-wavelength campaigns at lower energies were performed during 
the HESS data taking, consisting of several individual pointings with the  
XMM-Newton and RXTE satellites.} 
%
   {A strong VHE signal, $\sim$17$\sigma$ total, from PKS\,2005$-$489 
was detected during the four years of HESS observations (90.3 hrs live time). 
The integral flux above the average analysis threshold of 400 GeV is
$\sim$3\% of the flux observed from the Crab Nebula and varies weakly
on time scales from days to years.  The average VHE spectrum measured 
from $\sim$300 GeV to $\sim$5 TeV is characterized
by a power law with a photon index, $\Gamma = 3.20\pm0.16_{\rm stat}\pm0.10_{\rm syst}$.
At X-ray energies the flux is observed to vary by more than an order of magnitude 
between 2004 and 2005.
Strong changes in the X-ray spectrum ($\Delta$$\Gamma_{\mathrm X} \approx 0.7$) 
are also observed, which appear to be mirrored in the VHE band.}
   {The SED of PKS\,2005$-$489, constructed for the first time with contemporaneous data 
   on both humps,  shows significant evolution. 
   The large flux variations in the X-ray band, 
   coupled with weak or no variations
   in the VHE band and a similar spectral behavior, suggest the emergence 
   of a new, separate, harder emission component in September 2005.}
%
%

\authorrunning{The HESS Collaboration}
\titlerunning{VHE monitoring of PKS\,2005--489 and simultaneous multi-wavelength observations.}

   \maketitle

\section{Introduction}
PKS 2005$-$489 is one of the brightest BL Lac objects, 
at all wavelengths, in the Southern Hemisphere.
It was initially discovered as a strong radio source in the Parkes 2.7\,GHz 
survey \citep{discovery_paper} and later identified as a BL Lac object \citep{BL_id}. 
It belongs to the complete 1-Jy radio catalog \citep{stickel}, and 
is one of the few extragalactic objects detected in the EUV band \citep{euv}.
Its redshift, $z=0.071$ \citep{redshift}, is determined from weak, narrow emission 
lines observed during a low optical state.

PKS 2005$-$489 is classified as a High-frequency peaked BL Lac object \citep[HBL;][]{giommipadovani94},
because of its high X-ray-to-radio flux ratio \citep{Sambruna_paper}
and because its broad-band spectral energy distribution (SED) peaks
in the UV--soft X-ray band. As is typical of HBLs, the X-ray spectrum is dominated 
by the synchrotron emission of high-energy electrons.
The second SED hump is expected to peak in the GeV$-$TeV $\gamma$-ray band,
and is commonly believed to be produced 
 by the same electrons
up-scattering via the inverse Compton mechanism 
seed photons of lower energy.

In the X-ray band, PKS\,2005$-$489 has been studied extensively,
showing an extreme flux and spectral variability.
Large flux variations with correlated spectral hardening of 
the generally steep spectrum 
(photon index\footnote{The power-law spectrum is described as
$N(E)=N_0\,E^{-\Gamma}$} $\Gamma=2.7-3.1$) 
were observed during five EXOSAT observations \citep{giommi90,sambruna94}.
Two ROSAT observations in 1992 confirmed the EXOSAT results and similarly 
show a soft spectrum \citep[$\Gamma\simeq3$;][]{Sambruna_paper}.
A harder spectrum ($\Gamma=2.3$ from 2 to 10 keV) was observed 
in September 1996, during \sax observations of a brighter X-ray state \citep{padovani}.
In October$-$November 1998  PKS\,2005$-$489 underwent a period of exceptional activity,
with several strong X-ray flares.  RXTE monitoring observations \citep{perlman}
were performed during the entire epoch, and the 2$-$10 keV flux reached $\sim$3\ergs{-10},  
approximately 30 times higher than the ROSAT values.  These RXTE observations
yielded a detection up to 40 keV and showed variations in the photon index 
between $\Gamma=2.3$ and 2.8.  An X-ray flare alert also triggered \sax observations
on November 1$-$2, 1998.  The X-ray spectrum was measured between 0.1 to 200 keV
and was characterized by a curved shape and harder photon indices,
in both the soft ($\Gamma_1=2.0$) and hard ($\Gamma_2=2.2$) X-ray bands 
\citep[$<$ and $>$2 keV;][]{gt}.
More recently, observations  with the \swift satellite 
have generally shown PKS\,2005$-$489 
in a low-flux, steep-spectrum state \citep[$\Gamma\simeq3$,][]{massaro08}.

Strong correlations between X-ray and $\gamma$-ray emission
have been observed in many  HBL
\citep[e.g.][]{pian98,maraschi99,1959,fossati08,2155chandra},
and are typically expected in a synchrotron-Compton scenario.
Therefore the very bright flux and large flux/spectral variability 
at X-ray energies make PKS\,2005$-$489 one of the most promising targets
for observing a similar behavior in the $\gamma$-ray domain.

In the VHE ($E>100$ GeV) band, PKS\,2005$-$489 was detected neither during 
observations made between 1993 and 2000 by either the CANGAROO or 
Durham groups \citep{cangaroo1,cangaroo2,cangaroo3,durham},
nor by HESS with a partial array 
during its commissioning in 2003.  
In 2004, HESS discovered
VHE $\gamma$-ray emission from PKS\,2005$-$489 
with a significance of 6.7$\sigma$,
at a flux of a few percent of the Crab Nebula  \citep{hess_discovery}.  
The measured spectrum was soft ($\Gamma=4.0$). 
In the MeV-GeV band,
PKS\,2005$-$489 is one of the few HBL detected by  EGRET. 
However, the observed significance is marginal: 
3.7$\sigma$ above 100 MeV ~\citep{egret1} and 4.1$\sigma$ above 1 GeV \citep{egret2}.  
It was instead detected by {\it Fermi} with high significance ($>$10$\sigma$),
during the first three months of operation \citep[2008, Aug.--Oct.,][]{lbas}.

Because of the high potential for strong $\gamma$-ray activity, 
PKS\,2005$-$489 has been monitored at VHE by HESS every year since its detection in 2004.
Several campaigns of coordinated observations with the X-ray satellites \xmm~ and RXTE 
were also performed. These simultaneous  observations, sampling
the same particle distribution through two different emission processes, 
represent a powerful diagnostic tool for probing the conditions of the inner blazar jet, 
especially during flaring events \citep{coppi}. 
   \begin{figure}
   \centering
      \includegraphics[width=0.5\textwidth]{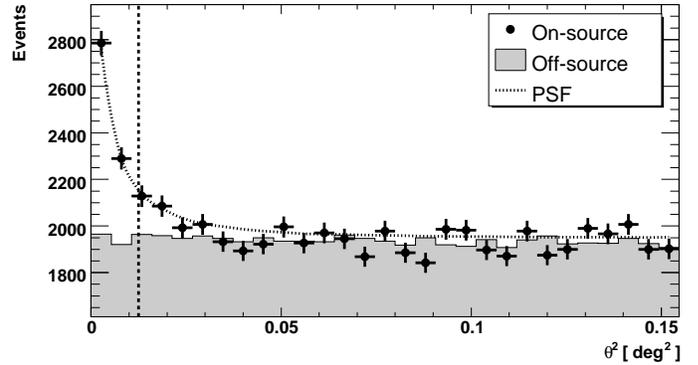}
      \caption{Distribution of $\theta^2$ for on-source events (points with statistical errors) and
        normalized off-source events (shaded) from observations
        of PKS\,2005$-$489.  The curve represents the $\theta^2$ distribution
	expected from simulations of a $\gamma$-ray point source (photon index $\Gamma=3.20$) at
	a zenith angle of $40^{\circ}$.
	The dashed line represents the cut on $\theta^2$ applied to the data.}
         \label{thtsq_plot}
   \end{figure}
  \begin{figure}
   \centering
      \includegraphics[width=0.5\textwidth]{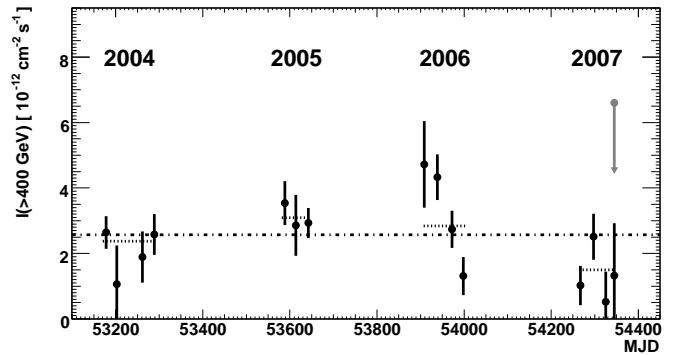}
      \caption{Integral flux, I($>$400 GeV), measured by HESS
from PKS\,2005$-$489 during each dark period of observations (i.e. moonless night time within a month).
Only the statistical errors are shown.
The horizontal line represents the average flux for all 
the HESS observations. The four horizontal line segments represent
the average annual flux observed during the corresponding years (see Table \ref{results}).}
         \label{monthly_plot}
   \end{figure}

In this article the results of all HESS observations taken 
from 2004 through 2007 are presented, together with the results of the  
multi-wavelength observations. These campaigns characterize the SED of PKS\,2005$-$489 
during different states and, for the first time, sample both humps of the SED simultaneously.
A re-analysis of the 2004 HESS data is also provided,
which benefits from an improved calibration of the absolute energy scale with respect 
to the previously published result.
    \begin{figure*}
   \centering
      \includegraphics[width=0.6\textwidth]{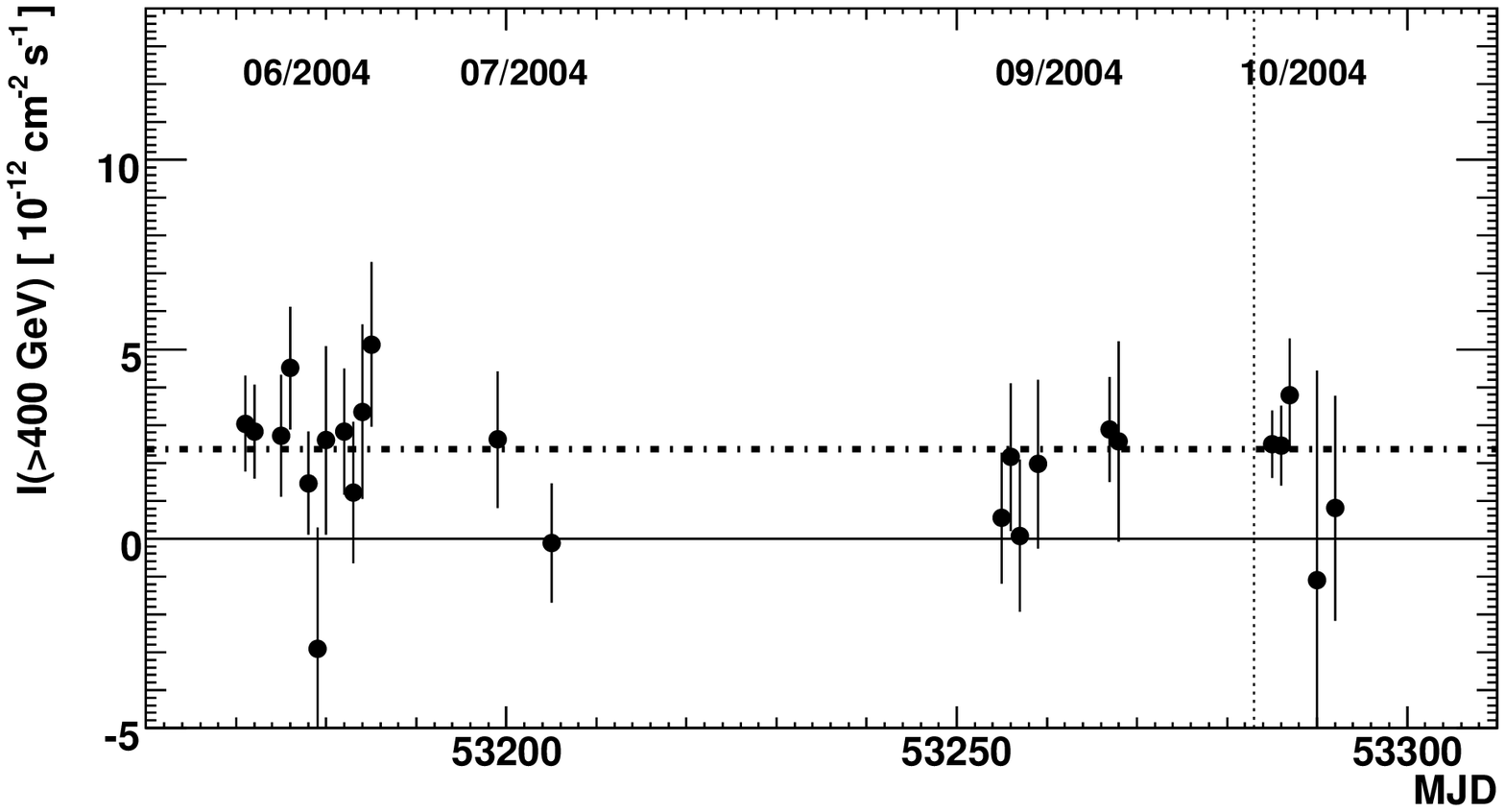}\\
      \includegraphics[width=0.6\textwidth]{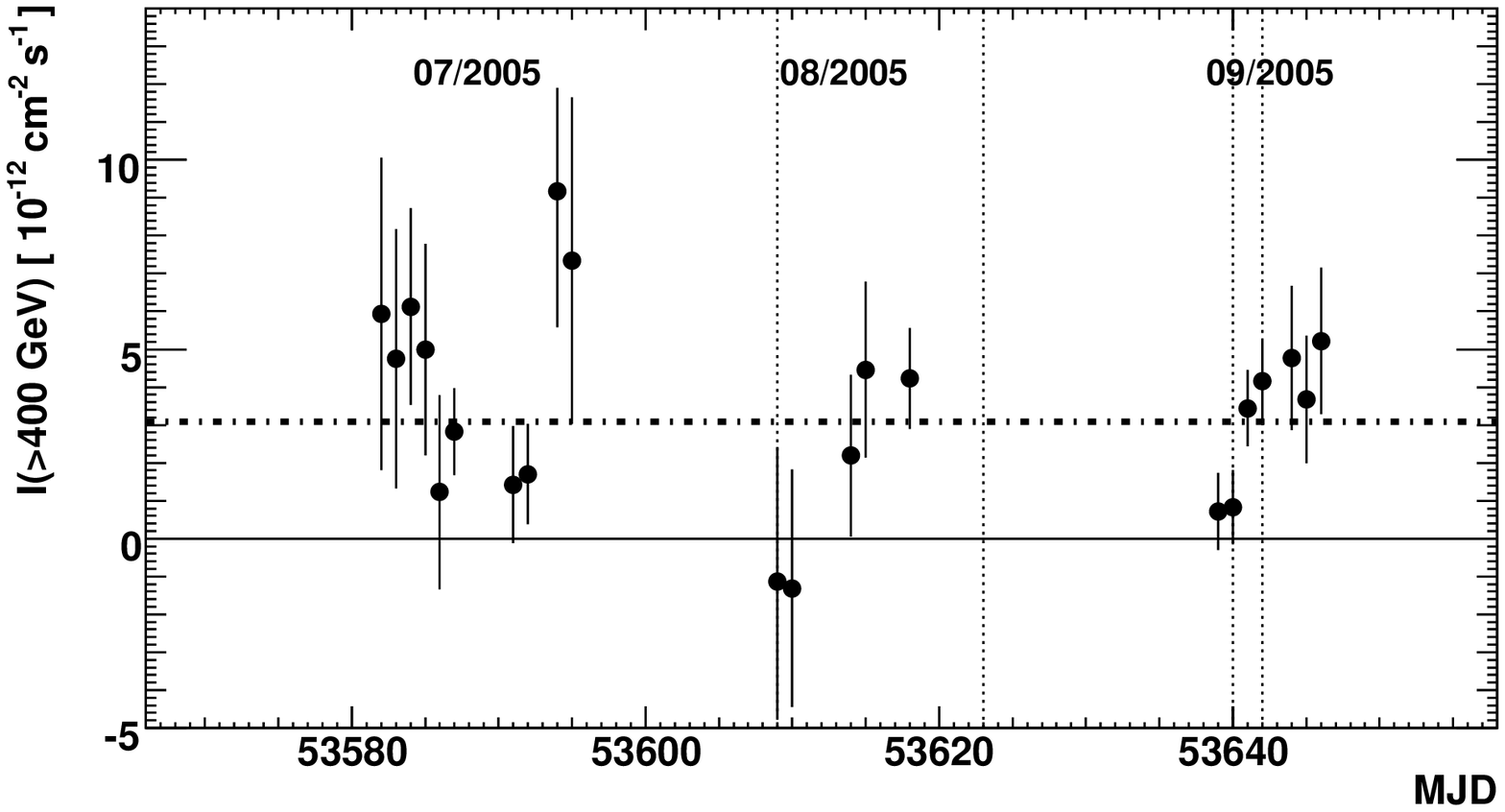}\\
      \includegraphics[width=0.6\textwidth]{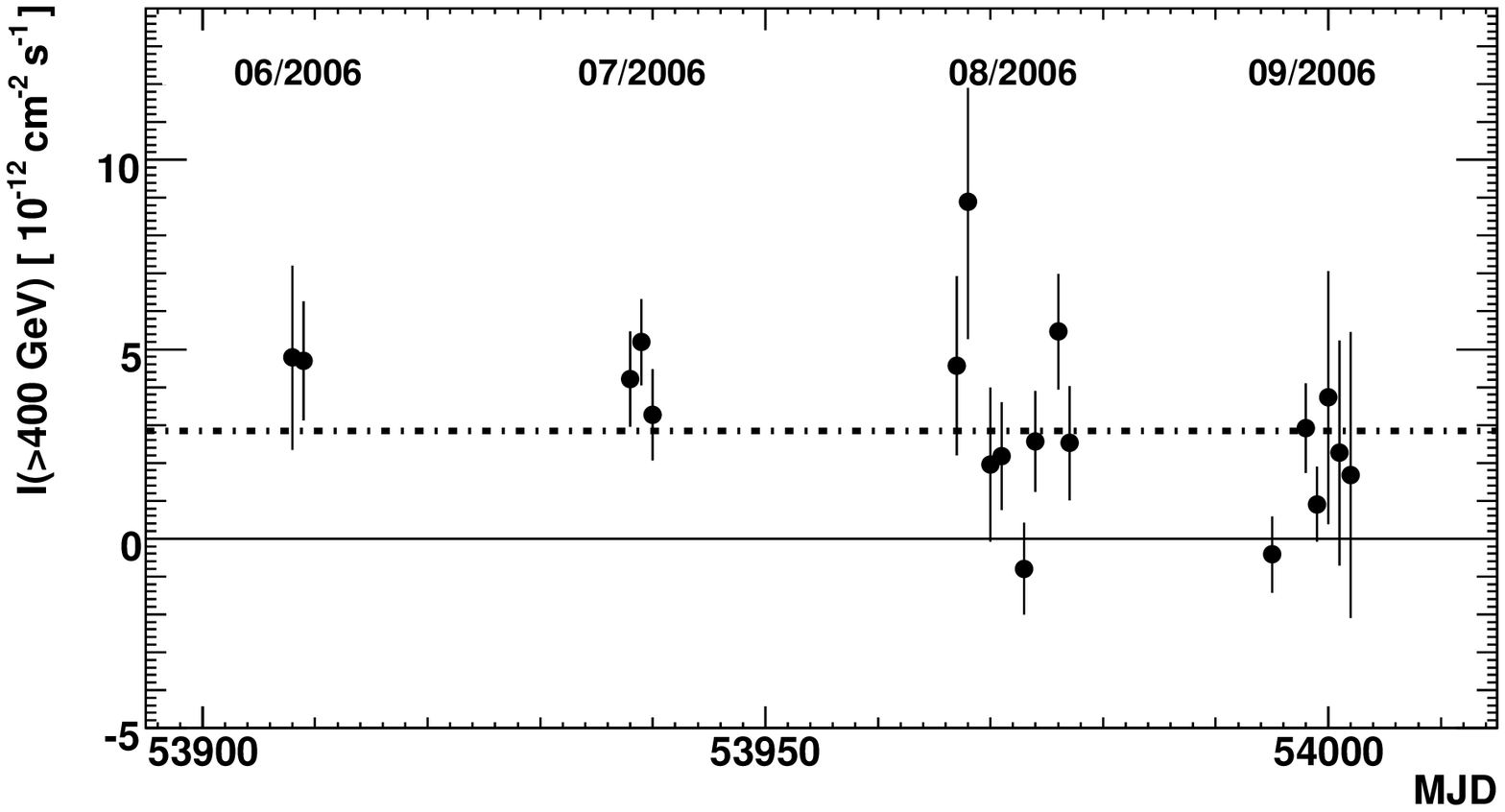}\\
      \includegraphics[width=0.6\textwidth]{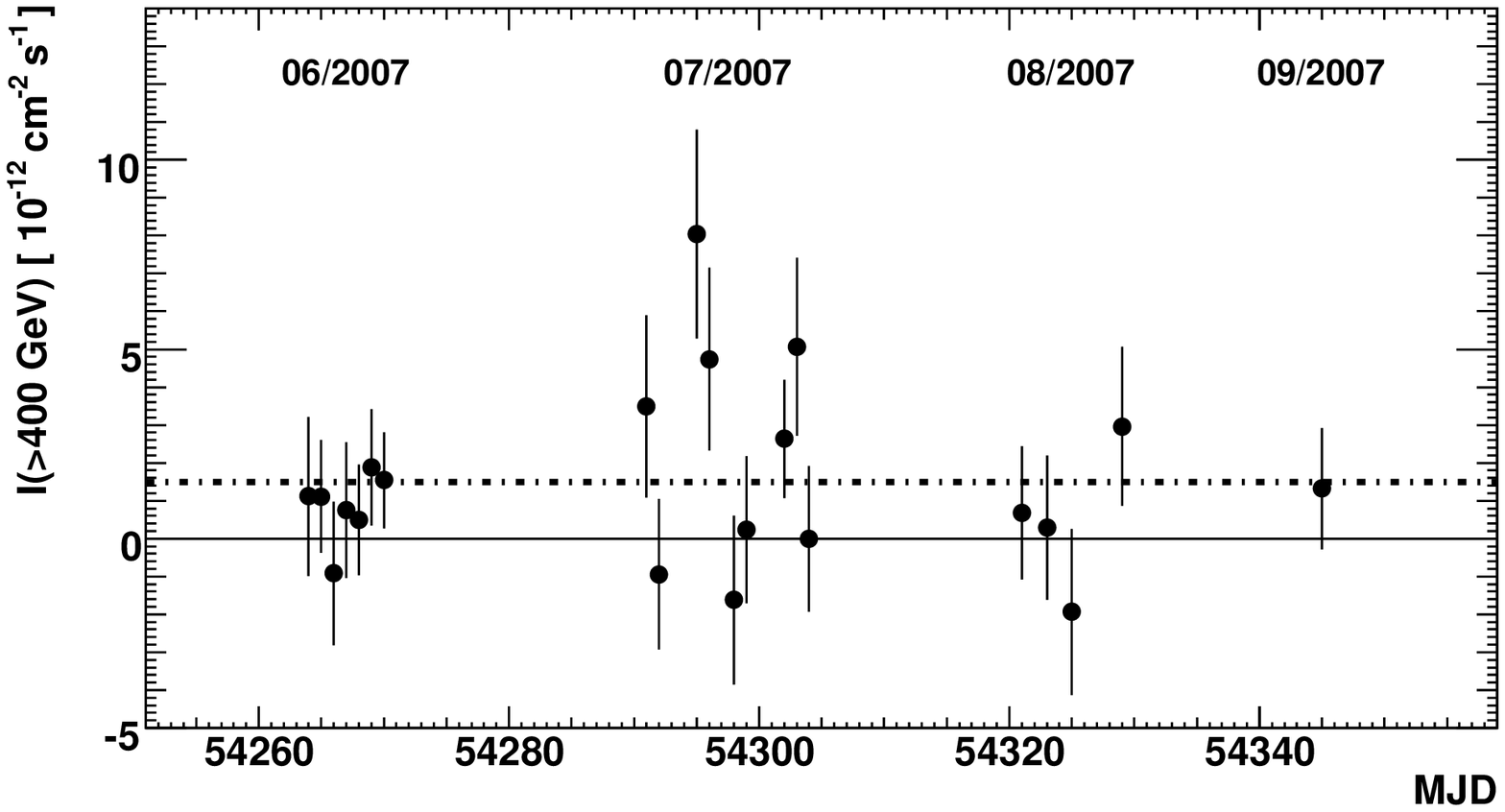}\\
      \caption{Integral flux, I($>$400 GeV), measured by HESS
from PKS\,2005$-$489 during each night of observations.  The 
individual plots represent each of the four years (2004$-$2007) of data taking.
Only the statistical errors are shown.  The horizontal lines represent
the average annual flux observed during the respective year.  The vertical
lines at MJD 53282, 53640, and 53642 represent the nights 
of \xmm~ observations. The epoch of the RXTE observations is
between the vertical lines at MJD 53609 and 53623.}
         \label{nightly_plots}
   \end{figure*}

\begin{table*}[t]
   \begin{minipage}[t]{2.0\columnwidth}
      \caption{Results from long-term HESS observations of PKS\,2005$-$489.}
         \label{results}
           \centering
         \begin{tabular}{c c c c c c c c c c c c c}
            \hline\hline
            \noalign{\smallskip}
            Dark & MJD & MJD & Time & On & Off & $\alpha$ & Excess & Sig 
& I($>$400 GeV)\footnote{The integral flux above 400 GeV is calculated using the excess of events above an energy threshold, which differs
slightly from the observed excess.  The quoted error is statistical only and the
20\% systematic error on the observed flux is not shown.} 
& Crab\footnote{The percentage is calculated relative to the HESS Crab Nebula flux above 400 GeV \citep{HESS_crab}.} 
& $\chi^2$\,\,(NDF\footnote{The $\chi^2$, degrees of freedom (NDF), and corresponding $\chi^2$ probability P($\chi^2$) are given
for fits of a constant to I($>$400 GeV) binned nightly within a dark period, or monthly within a year, 
or yearly within the total.}) 
& P($\chi^2$)$^c$\\
            Period & First & Last & [hrs] & & & & &  [$\sigma$] & [10$^{-12}$\,cm$^{-2}$\,s$^{-1}$] & \% & & \\
            \noalign{\smallskip}
            \hline
            \noalign{\smallskip}
            06/2004 & 53171 & 53185 & 8.1  & 678 & 5877  & 0.0919 & 138 & 5.4 & $2.64\pm0.50$ & 3.0 & 7.2\,\,(10) & 0.71\\
            07/2004 & 53199 & 53205 & 1.5  & 105 & 985   & 0.0909 & 15  & 1.5 & $1.06\pm1.18$ & 1.2 & 1.4\,\,(1)  & 0.75\\
            09/2004 & 53255 & 53268 & 5.6  & 342 & 3171  & 0.0916 & 52  & 2.8 & $1.89\pm0.78$ & 2.1 & 2.1\,\,(5)  & 0.84\\
            10/2004 & 53285 & 53292 & 9.0  & 569 & 5065  & 0.0916 & 105 & 4.5 & $2.58\pm0.62$ & 2.9 & 1.5\,\,(4)  & 0.83\\
            07/2005 & 53582 & 53595 & 9.4  & 573 & 4504  & 0.0908 & 164 & 7.3 & $3.56\pm0.67$ & 4.0 & 9.2\,\,(9)  & 0.42\\
            08/2005 & 53609 & 53618 & 5.1  & 286 & 2159  & 0.0989 & 73  & 4.5 & $2.86\pm0.93$ & 3.2 & 4.6\,\,(4)  & 0.33\\
            09/2005 & 53639 & 53646 & 18.1 & 1072 & 9125 & 0.0928 & 225 & 7.1 & $2.93\pm0.46$ & 3.3 & 13.1\,\,(6) & 0.041\\
            06/2006 & 53908 & 53909 & 1.3  & 127 & 824   & 0.0938 & 50  & 4.9 & $4.72\pm1.32$ & 5.3 & 0.0\,\,(1)  & 0.97\\
            07/2006 & 53938 & 53940 & 4.4  & 397 & 2927  & 0.0901 & 133 & 7.3 & $4.33\pm0.70$ & 4.8 & 1.3\,\,(2)  & 0.52\\
            08/2006 & 53967 & 53977 & 8.4  & 500 & 4261  & 0.0913 & 111 & 5.1 & $2.74\pm0.57$ & 3.1 & 15.1\,\,(7) & 0.035\\
            09/2006 & 53995 & 54002 & 7.4  & 405 & 4116  & 0.0920 & 26  & 1.3 & $1.31\pm0.58$ & 1.5 & 6.0\,\,(5)  & 0.30\\
	    06/2007 & 54264 & 54270 & 5.3 & 333 & 3006 & 0.0924 & 55 & 3.1 & $1.02\pm0.60$ & 1.1 & 1.9\,\,(6) & 0.93\\
	    07/2007 & 54291 & 54304 & 4.4 & 309 & 2412 & 0.0964 & 76 & 4.5 & $2.51\pm0.70$ & 2.8 & 15.2\,\,(8) & 0.056\\
	    08/2007 & 54321 & 54329 & 1.8 & 93 & 849 & 0.0977 & 10 & 1.0 & $0.52\pm0.92$ & 0.6 & 2.9\,\,(3) & 0.41\\
	    09/2007 & 54345 & 54345 & 0.5 & 11 & 114 & 0.1000 & 0 & $-0.1$ & $<6.60$\footnote{The upper limit is calculated
at a 99.9\% confidence level \citep{UL_tech}.} & $<$7.4 & $-$ & $-$ \\
            \noalign{\smallskip}
            \hline
            \noalign{\smallskip}
            2004 & 53171 & 53292 & 24.2 & 1694 & 15098 & 0.0917 & 310 & 7.7 & $2.37\pm0.33$ & 2.6 & 2.0\,\,(3) & 0.57\\
            2005 & 53582 & 53646 & 32.6 & 1931 & 15785 & 0.0930 & 462 & 11.0 & $3.09\pm0.35$ & 3.5 & 0.6\,\,(2) & 0.73\\
            2006 & 53908 & 54002 & 21.5 & 1429 & 12128 & 0.0914 & 320 & 8.8 & $2.84\pm0.34$ & 3.2 & 13.5\,\,(3) & 0.0037 \\
	    2007 & 54264 & 54345 & 12.0 & 746 & 6381 & 0.0947 & 141 & 5.3 & $1.50\pm0.40$ & 1.7 &  3.8\,\,(3) & 0.28\\
            \noalign{\smallskip}
            \hline
            \noalign{\smallskip}
             Total & 53171 & 54345 & 90.3 & 5800 & 49392 & 0.0924 & 1233 & 16.7 & $2.57\pm0.18$ & 2.9 & 10.2\,\,(3) & 0.016 \\
            \noalign{\smallskip}
            \hline
       \end{tabular}
    \end{minipage}
   \end{table*}

\section{HESS observations and analysis technique}
PKS\,2005$-$489 was observed with the HESS array \citep[][]{hess} 
for a total of 158.0 hours 
(352 runs of $\sim$28 min each) from 2004 through 2007.  During these observations
the array tracked a position offset from the blazar by 0.5$^{\circ}$
in alternating directions to
enable both on-source observations and simultaneous estimation
of the background induced by charged cosmic rays.
A total of 207 runs pass the standard HESS data-quality 
selection, yielding an exposure of 90.3 hrs live time at 
a mean zenith angle $Z_{\rm mean} = 35^{\circ}$.
The results presented here were generated using 
the standard HESS calibration methods \citep{calib_paper}
and analysis tools \citep{std_analysis}, with the
{\it standard cuts} event-selection criteria
(except for the 2007 spectrum, see Sect 3.2).
On-source data were taken from a circular region
of radius $\theta_{cut}$=0.11$^{\circ}$ centered
on PKS\,2005$-$489, and the background (off-source data) was estimated using
the {\it Reflected-Region} method \citep{bgmodel_paper}.
Equation (17) in \citet{lima} was used to calculate
the significance of any excess.   All VHE integral fluxes 
reported throughout this article were calculated assuming 
the time-average photon index of $\Gamma=3.20$ determined
in Sect.~\ref{Spectrum_study}.

The PKS\,2005$-$489 observations presented here span
four years (2004$-$2007) of HESS data taking.  During this time
the optical throughput of the instrument decreased,
because of the degradation of the reflective surfaces
of the mirrors and Winston cones, as
well as  accumulation of dust on the optical elements.
For the entire data sample,
the optical efficiency has decreased by an average of  
28\% compared to a newly commissioned instrument,
with its mirrors installed in Oct. 2001, Dec. 2002, June 2003, 
and August 2003 on CT3, CT2, CT4, and CT1, respectively.
To minimize the effects of long-term variation in
the optical efficiency of the HESS array, the estimated
energy of each event was corrected
using the ratio of efficiencies determined on a run-wise basis
from simulated and observed muons \citep{HESS_crab}. After accounting for 
the decreasing optical throughput of the 
HESS array, the average energy threshold of the 
analysis at $Z_{\rm mean}$ is 400 GeV.

\section{HESS results}
PKS\,2005$-$489 was clearly detected in each of the four years (2004$-$2007)
that it was observed by HESS.  A total of 1233 excess events, corresponding
to a statistical significance of 16.7 standard deviations ($\sigma$),
was detected from the direction of  the blazar.  The results
of the HESS observations are given in Table~\ref{results}, which shows
the dead time corrected observation time, 
the number of on and off-source events, the on/off normalization ($\alpha$),
the excess and the corresponding statistical significance, for various temporal
breakdowns of the HESS data sample.
   \begin{table*}
     \begin{minipage}[t]{2.0\columnwidth}
      \caption{Best $\chi^2$ fits of a power-law to the various
	spectra of PKS\,2005$-$489 measured by HESS.}  
         \label{annual_results}
        \centering
         \begin{tabular}{c c c c c c c c}
            \hline\hline
            \noalign{\smallskip}
             Epoch & E$_{\rm lower}$ & E$_{\rm upper}$ & $\Gamma$ & $I_{400}$ & $\chi^2$ & NDF & P($\chi^2$)\\
		& [TeV] & [TeV] & & [$10^{-11}$ cm$^{-2}$\,s$^{-1}$\,TeV$^{-1}$] & & & \\
            \noalign{\smallskip}
            \hline
            \noalign{\smallskip}
	    2004 (AH05)\footnote{This (AH05) is the previously published HESS result \citep{hess_discovery}
	for 2004, and was not corrected for long-term changes in the optical efficiency of the system.} & 
0.2 & 2.7 & $3.98\pm0.38_{\rm stat}\pm0.10_{\rm syst}$ & $0.74\pm0.27_{\rm stat}\pm0.15_{\rm syst}$ & 5.6 & 7 & 0.59 \\
            \noalign{\smallskip}
            \hline
            \noalign{\smallskip}
	    2004\footnote{The 2004 entry is the very same data as presented in AH05,
	but with a correction (see text) applied to account for relative changes
in the optical throughput of the HESS array. The 2005, 2006, 2007, and
	total entries also have this correction applied.} & 0.30 & 4.0 & $3.65\pm0.39_{\rm stat}\pm0.10_{\rm syst}$ & $1.62\pm0.33_{\rm stat}\pm0.32_{\rm syst}$ & 6.7 & 7 & 0.46 \\
            2005 & 0.32 & 5.1 & $3.09\pm0.22_{\rm stat}\pm0.10_{\rm syst}$ & $1.46\pm0.28_{\rm stat}\pm0.29_{\rm syst}$ & 13.3 & 6 & 0.038\\
            2006 & 0.35 & 11 & $2.86\pm0.20_{\rm stat}\pm0.10_{\rm syst}$ & $1.28\pm0.20_{\rm stat}\pm0.26_{\rm syst}$ & 3.7 & 7 &  0.53\\
 	    2007\footnote{The 2007 spectrum is generated with the {\it spectrum cuts} 
\citep{HESS_1553}, see Sect 3.2} 
& 0.24 & 0.75 & $3.48\pm0.68_{\rm stat}\pm0.10_{\rm syst}$ & $1.00\pm0.22_{\rm stat}\pm0.20_{\rm syst}$ & 0.6 & 2 &  0.74\\
            \noalign{\smallskip}
            \hline
            \noalign{\smallskip}
	    Total & 0.3 & 5.3 & $3.20\pm0.16_{\rm stat}\pm0.10_{\rm syst}$ & $1.37\pm0.17_{\rm stat}\pm0.27_{\rm syst}$ & 9.9 & 8 & 0.27\\
          \noalign{\smallskip}
            \hline
       \end{tabular}
     \end{minipage}
   \end{table*}

Figure~\ref{thtsq_plot} shows the on-source and normalized off-source
distributions of the square of the angular difference between
the reconstructed shower position 
and the source position ($\theta^{2}$) for all observations. 
As can be seen in the figure, the distribution 
of the excess (i.e. the observed signal) is very similar
to what is expected from a simulated point-source of VHE $\gamma$-rays 
at comparable zenith angles.  The off-source distribution 
is approximately flat in $\theta^{2}$, as expected. 

The map of excess counts is well-fit by a two-dimensional 
Gaussian with a centroid located at
$\alpha_{\rm J2000}=20^{\mathrm h}09^{\mathrm m}27.0^{\mathrm s}\pm1.5^{\mathrm s}_{\rm stat}\pm1.3^{\mathrm s}_{\rm syst}$
and $\delta_{\rm J2000}=-48^{\circ}49'52''\pm16''_{\rm stat}\pm$$20''_{\rm syst}$.
As expected, the fit location of HESS\,J2009$-$488 is consistent\footnote{The 
difference between
the two positions is $25''\pm28''$.} with the position 
($\alpha_{\rm J2000}=20^{\mathrm h}9^{\mathrm m}25.4^{\mathrm s}$, 
$\delta_{\rm J2000}=-48^{\circ}49'54''$)
of the blazar~\citep{position}.  
The upper limit (99\% confidence level)
on the extension of HESS\,J2009$-$488 is 1.0'.

\subsection{VHE Flux}
The observed integral flux above 400 GeV for the entire data set is
I($>$400 GeV) = $(2.57\pm0.18_{\rm stat}\pm0.51_{\rm syst}) \times 10^{-12}$ 
cm$^{-2}$\,s$^{-1}$.  This corresponds to $\sim$2.9\% of 
the flux above 400 GeV from the Crab Nebula, as determined by HESS \citep{HESS_crab}.
Figures ~\ref{monthly_plot} and ~\ref{nightly_plots} 
show the flux measured for each dark period and night, respectively.
 
The integral flux I($>$400 GeV) observed
during various epochs is reported in Table~\ref{results}, 
together with the $\chi^2$ and corresponding probability, P($\chi^2$),
for fits of a constant to the data when binned by nights within each dark period, by dark periods within
a year, and by year within the total observations.  There are weak indications 
(P($\chi^2) < 0.05$) of variability on annual time scales, monthly time scales in 2006,
and nightly time scales in the dark periods of September 2005 and August 2006.
The combined variations on different time scales lead to an indication
of variability in the overall monthly curve 
($\chi^2=30.5$ for 13 NDF, with P($\chi^2$)$\lesssim0.004$, Fig. ~\ref{monthly_plot}).
Although no variability is seen inside the other epochs (e.g. monthly time scales in 2004, 2005,
or 2007), variations in amplitude comparable to the statistical errors
cannot be ruled out.

\subsection{VHE spectra \label{Spectrum_study}}
\begin{figure}[t]
   \centering
      \includegraphics[width=0.5\textwidth]{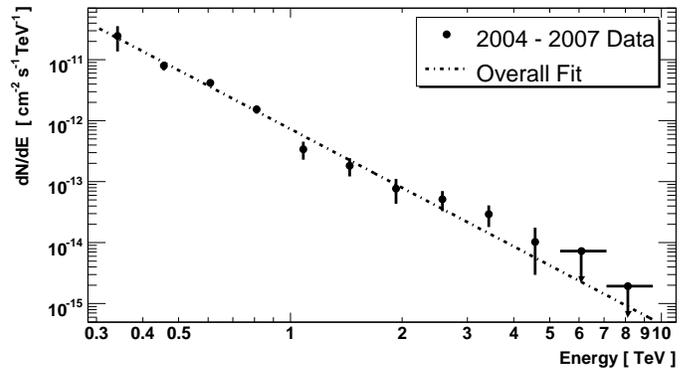}
      \caption{Time-average VHE energy spectrum observed from PKS\,2005$-$489. 
	The dashed line represents the best $\chi^2$ fit of a power law to
        the observed data (see Table 2), and then extrapolated to 10 TeV.  
	Only the statistical errors are shown and 
	the upper limits are at the 99\% confidence level \citep{UL_tech}.}
         \label{spectrum_plot}
   \end{figure}

The time-average photon spectrum for the entire data set is shown 
in Fig.~\ref{spectrum_plot}. All points in the energy range 
300 GeV to $\sim$5.3 TeV are significant except the 
last (1.5$\sigma$) at $\sim$4.6 TeV.  The data are well-fit,
$\chi^2$ of 9.9 for 8 degrees of freedom,
by a power law (dN/dE = $I_{400}$ (E / 400 GeV)$^{-\Gamma}$)
with a photon index 
$\Gamma=3.20\pm0.16_{\rm stat}\pm0.10_{\rm syst}$.
Removing the $\sim$4.6 TeV point does not significantly
affect the fit result. No evidence is found 
for significant features, such as a cut-off or break, in the energy spectrum,
also considering the upper limits at higher energies.

\begin{figure*}[t]
   \centering
      \includegraphics[width=1.0\textwidth]{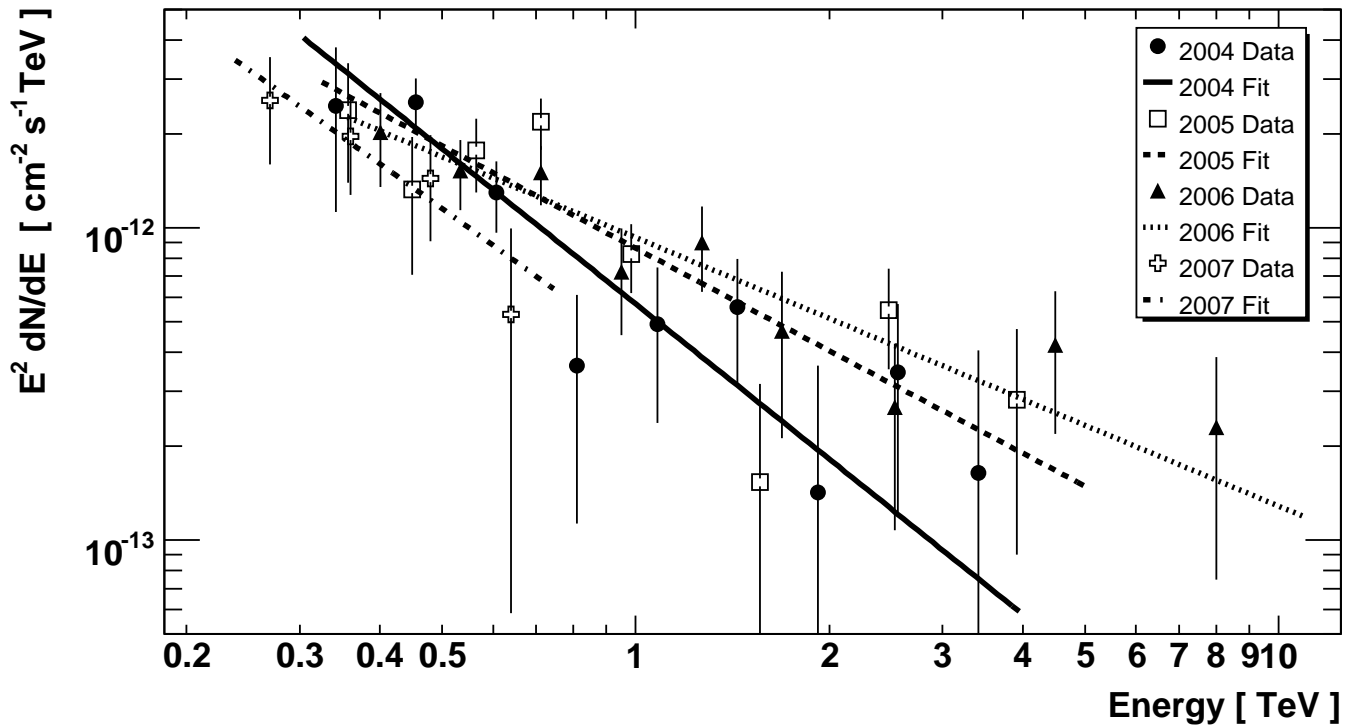}
      \caption{Photon spectra observed by HESS 
	from PKS\,2005$-$489 during each year of data taking. 
	 Only the statistical errors are shown.
	Each line represents the best $\chi^2$ fit of a power law to
        the observed data.}
         \label{annual_spectra}
   \end{figure*}

The time-average spectra measured during each year of data-taking are shown
in Fig.~\ref{annual_spectra}.  The results of the best $\chi^2$ fits
of a power law to these data are shown in Table~\ref{annual_results}.  
All spectra are generated with {\it standard cuts} \citep{std_analysis}, 
except for 2007.  Given the 
 short exposure, low flux, steep spectral slope,  and degradation
of the optical efficiency, a spectrum for 2007 could only be generated 
with the {\it spectrum cuts} \citep{HESS_1553}, 
which lower the energy threshold of the analysis at the expense
of sensitivity at higher energies. 
In Table 2, the epoch, lower and upper energy bounds, photon index, differential flux 
normalization at 400 GeV ($I_{400}$), $\chi^2$, 
degrees of freedom (NDF), and $\chi^2$ probability P($\chi^2$) 
for each fit are given.
The $\chi^2$ probability for a fit of a constant 
to the annual $\Gamma$ values is 0.30.

\subsection{Correction to 2004 HESS data}
\begin{figure}
\centering
   \includegraphics[width=0.5\textwidth]{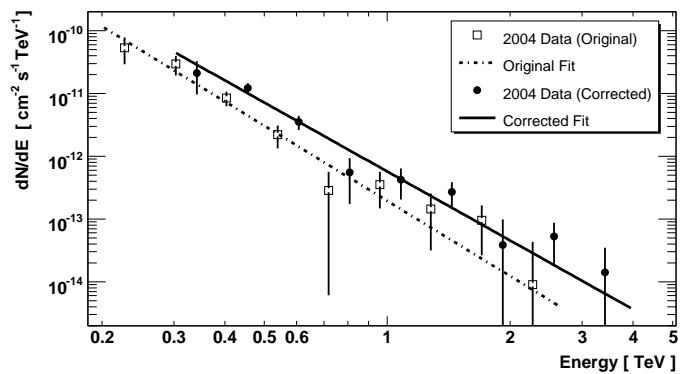}
   \caption{VHE spectrum measured by HESS from PKS\,2005$-$489 in
     2004 compared to the previously published version 
     \citep{hess_discovery}.	     Only the statistical errors are shown.
     Correcting the 2004 data for decreases in the 
     optical efficiency of the HESS array results 
     in a three-times higher integral flux above 400 GeV.}
      \label{muoncorr_plot}
\end{figure}

The data previously published \citep{hess_discovery}
for HESS observations of PKS\,2005$-$489 in 2004 were
not corrected for a 19\% average decrease 
in the optical efficiency until 2004. 
In addition, the early simulations neglected the small gaps between
Winston cone apertures, the combined effect being an overall correction
of 22\% compared to the initial analysis.
Figure~\ref{muoncorr_plot} illustrates
the effect of correcting the energy of individual events for the relative 
optical efficiency of the system, as described earlier.    
Comparing (see Table~\ref{annual_results})
the spectrum for 2004 determined here to 
the previously published one shows
significant differences in the flux normalization 
($I_{400}$) but not in the photon index ($\Gamma$). 
The flux measured in 2004 is three times
higher than previously published\footnote{The flux was reported
above a threshold of 200 GeV.  Extrapolating the earlier result, 
using $\Gamma=4.0$ as reported, yields 
I($>$400 GeV) = $(0.86\pm0.12_{\rm stat}\pm0.17_{\rm syst}) \times 10^{-12}$ 
cm$^{-2}$\,s$^{-1}$.}, because of the 
steep spectrum of the source. 
It should be noted that, according to the model for the decrease 
in the optical efficiency of the HESS system, the simulated values
match the actual values measured in 2003, therefore the upper limit
published for 2003 HESS observations is unchanged.  Extrapolating
the 2003 upper limit 
\citep[99\% confidence level,][]{hess_discovery}
to above 400 GeV (using $\Gamma=4.0$ from the original publication) yields
I($>$400 GeV) $< (0.65\pm0.13_{\rm syst}) \times 10^{-12}$ cm$^{-2}$\,s$^{-1}$.

\section{Coordinated XMM-Newton/RXTE observations}
PKS\,2005$-$489 was observed three times with \xmm~ between 2004 and 2005
(Obs-Id 0205920401, 0304080301 and 0304080401).
The satellite pointings were scheduled such that simultaneous HESS
observations were possible.  The first \xmm~ pointing 
was performed on October 4, 2004, and yielded a net exposure of 11 ks.
Unfortunately poor weather did not allow HESS observations on this night.
Good quality HESS data were taken on the following nights.  
Two more pre-planned \xmm~ observations were performed on the nights of 
September 26 and 28, 2005,
with exposures of 21 and 25 ks, respectively. The goal
of scheduling such temporally close\footnote{The blazar was observed in 2 consecutive orbits.} 
\xmm~ pointings was to sample both the shortest variability time scales
and  spectral variations occurring on the typical time scale of HBL 
\citep[one to a few days; see e.g.][]{tanihata}.  It should also be noted that the
simultaneous observations had to occur near the end of the HESS observing 
season for PKS\,2005$-$489, because of the narrow overlap between the 
HESS and \xmm~ visibility windows caused by the 
constraints on the orientation of the solar panels of the satellite.

\begin{figure*}[th]
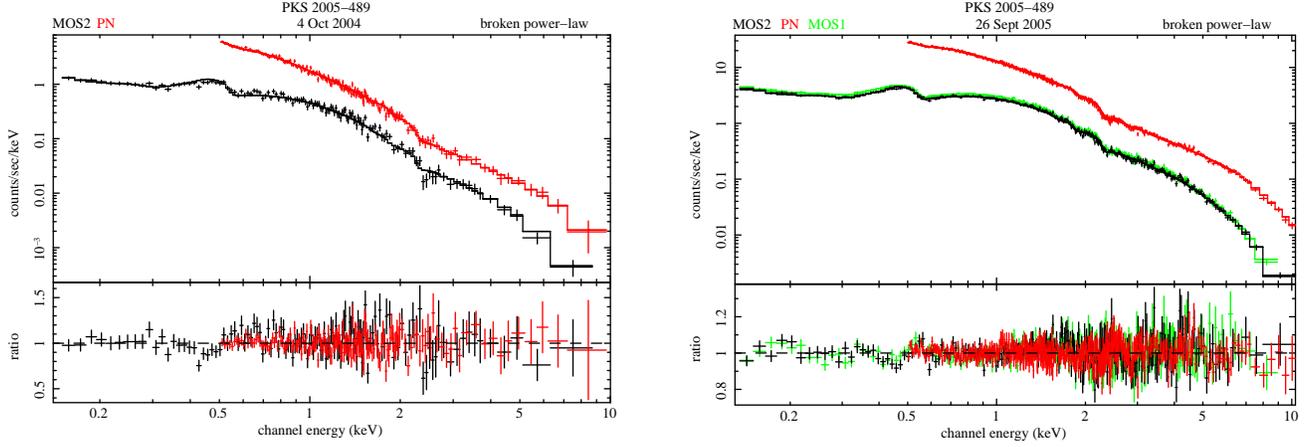

\centering
    \includegraphics[angle=-90,width=0.45\textwidth]{figures/xmm2004c.ps}\hspace{0.7cm}
    \includegraphics[angle=-90,width=0.45\textwidth]{figures/xmm2005c.ps}
   \caption{Best-fit data and folded model, plus residuals, 
   of the \xmm~ PN (upper) and MOS2/1 (lower) spectra of PKS\,2005$-$489, 
   in October 4, 2004 (left panel) and September 26, 2005 (right panel).
   The spectra are fitted with a concave broken power-law model plus Galactic absorption.
}
\label{xplots}
\end{figure*}

\begin{table*}
\caption{Fits to the X-ray spectra$^{a}$ in the different epochs.}
\hspace{-0.2cm}
\label{xfits}
\begin{tabular}{lccccccccc}
\hline
\hline
Instrument  & Band &  Date &  Model$^{b}$ & $\Gamma_1$ & $E_{\rm break}$    & $\Gamma_2$ 
&$F_{0.1-2.0~\rm keV}^{b}$  & $F_{2-10~\rm keV}^{b}$ & $\chi^2_r$\,(NDF) \\
\vspace*{1mm}
          &  keV &      &        &           & keV             &          &  \multicolumn{2}{c}{ergs cm$^{-2}$s$^{-1}$} &  \\
\vspace*{-3mm}\\
\hline
\vspace*{-3mm} \\ 
{\tiny XMM Mos2+PN}& {\tiny 0.15--10}  & {\tiny 4/10/04}       & p.l. & $3.04\pm0.02$ & --    &  --		      & 2.46\e{-11} & 1.03\e{-12} & 1.110\,(284) \\
            &     &          & b.p.l. & $3.06\pm0.02$ & $2.5\pm0.5$ &  $ 2.64\pm0.20$ & 2.77\e{-11} & 1.19\e{-12} & 1.048\,(282)     
\vspace*{1mm}\\  
{\tiny RXTE PCU2} & {\tiny 3--20} & {\tiny 20/8-19/9/05}  & p.l. & -- & -- &   $2.90\pm0.18$ & -- & 7.54\e{-12} & 0.79\,(16)
\vspace*{1mm}\\  
{\tiny XMM Mos+PN}&{\tiny 0.15--10} & {\tiny 26/9/05}      & p.l. & $2.31\pm0.01$ & --  &  --  & 7.55\e{-11} & 1.93\e{-11} & 1.138\,(860)    \\
             &     &         & b.p.l. & $2.34\pm0.01$ & $0.79\pm0.11$ &  $ 2.30\pm0.01$ & 7.72\e{-11} & 1.95\e{-11} & 1.122\,(858)
\vspace*{1mm}\\  
{\tiny XMM Mos+PN}&{\tiny 0.15--10}  &{\tiny  28/9/05}     & p.l. & $2.37\pm0.01$ & -- &  -- & 8.01\e{-11} & 1.77\e{-11} & 1.509\,(888)  \\
           &       &         & b.p.l. & $2.39\pm0.01$ & $1.78\pm0.15$ &  $ 2.31\pm0.02$ & 8.19\e{-11} & 1.83\e{-11} &  1.372\,(886) 
\vspace*{1mm} \\
\hline
\end{tabular}
\scriptsize
\vspace{-0.2cm}
 \begin{list}{}{}
 \item[$^a$] The errors are reported at a 90\% confidence level for 1 parameter ($\Delta\chi^2=2.71$).
 \item[$^b$] Fits with fixed Galactic $N_{\rm H}=3.93\times10^{20}$ cm$^{-2}$; power-law (p.l.) and broken power-law (b.p.l.) models. Unabsorbed fluxes. 
\end{list}
\end{table*}

During the \xmm~ observations, all three EPIC CCD cameras were used along 
with the thin filter. The PN detector \citep{pn} was operated in {\it timing mode},
i.e. when the data from a predefined area of the CCD chip are collapsed into a one-dimensional
row to be read out at high speed.
This mode allows the sampling of the shortest possible flux and spectral variations 
without potential pile-up\footnote{That is, when more than one X-ray photon
arrives in one camera pixel or adjacent pixels before the CCD is read out.} 
problems during bright states.
The two MOS \citep{mos} instruments were used in different configurations.
MOS1 took data in {\it timing mode} in 2004 and in {\it small-window mode} during both 2005
observations.  The MOS2 camera was  operated in {\it small-window mode} in 2004
and  in {\it large-window mode} in 2005.
Simultaneous observations were also performed with the Optical Monitor (OM) onboard \xmm,
using all photometric filters sequentially.  For these data the central window was read
in {\it fast mode} to enable temporal studies within the data. The exposure for each filter 
varied between 1800 and 4400 s, ultimately constrained by the overall duration of the pointing.

In August 2005, target-of-opportunity observations of PKS\,2005$-$489 with RXTE were 
triggered based on apparent enhanced VHE activity in preliminary analysis
of uncalibrated HESS data.  The satellite pointings were scheduled such that further
HESS data could be taken simultaneously.  Unfortunately, only a limited subset 
of the HESS data in this epoch passes standard quality selection criteria 
because of poor weather conditions in Namibia.
%
%

\subsection{XMM-Newton data analysis}
The EPIC data were processed and analyzed with {\tt SAS v7.1.0},
using the calibration files as of July 2008 and 
standard screening criteria\footnote{see SAS User Guide and {\tt CAL-TN-0018}.}.
A period of $\sim$30 minutes at the end of Obs-ID 
0304080401 was excluded from the analysis because of high background rates.
The MOS1 and MOS2 imaging observations are slightly to moderately affected by pile-up.
These effects were explored using {\tt epatplot} (which computes the fractions of single
and double pixel events showing if they differ from normal values), 
and are sufficiently suppressed by
the use of only single-pixel (PATTERN=0) events in the analysis.  Background 
events for MOS2  are extracted from an annulus around the source with 
inner radius $125\arcsec$ and outer radius $160\arcsec$. 
For MOS1 the background information was taken from
the same source region in blank-sky observations, since there is no region free 
of source photons on the same CCD.  A check was also performed using background 
regions from  the outer CCDs, as suggested in the SAS Analysis Guide. 
For the PN, 
the spectra were extracted from a rectangular box 20 pixels wide,
centered on the source strip and extended along the CCD 
(in raw pixel coordinates, 28$\leq$\verb+RAWX+$\leq$48).  The background events 
were  extracted from rows 2$\leq$\verb+RAWX+$\leq$18, avoiding 
the noisier strip on the CCD border.

The spectral analysis was performed with  {\tt XSPEC v11.3.2ag},
testing different binning schemes with at least 50 counts in each new bin.
For the 2004 observation, the spectrum from the MOS1 instrument is not included
since the {\it timing mode} data are significantly noisier than, but consistent with, 
the other detectors.
The MOS  and PN spectra are fit together, with a free constant
to allow for different MOS/PN normalizations. The PN flux is adopted as the
reference value, but the MOS fluxes are typically within a few percent of the PN fluxes.
The X-ray spectra are each fit with single and broken power-law models,
with interstellar absorption modelled by {\tt TBabs} \citep{wilms}.
This model is used with the solar abundances of \citet{wilms} 
and the cross-sections by \citet{verner96} \citep[for a discussion, see][]{baum06}.
 The results of the fits are shown in Fig. \ref{xplots} and Table \ref{xfits} 
(details in Sect. 5.2).

\begin{table}[t]
\caption[]{Optical Monitor parameters and source fluxes for the three \xmm~ observations.}
\label{om}
\scriptsize
\begin{tabular}{lcccccc}
\hline
\hline
\noalign{\smallskip}
Filter & $\lambda_{e}^{\mathrm{a}}$ & $A_{\lambda_{e}}/A_{\rm V}^{\mathrm{b}}$ & $F_{conv}^{\mathrm{c}}$  &
$F_{4/10}^{\mathrm{d}}$  &  $F_{26/9}^{\mathrm{d}}$  & $F_{28/9}^{\mathrm{d}}$  \\
\noalign{\smallskip}
\hline
\noalign{\smallskip}
$V$    & 5430 & 1.016 & 2.49  $\times 10^{-16}$ & 11.84  & 14.51 &  14.42  \\
$B$    & 4500 & 1.293 & 1.29  $\times 10^{-16}$ & 9.64   & 12.08 &  12.09  \\
$U$    & 3440 & 1.634 & 1.94  $\times 10^{-16}$ & 7.48   & 9.79  &  9.78  \\
$UVW1$ & 2910 & 1.870 & 4.76  $\times 10^{-16}$ & 6.12   & 8.27  &  8.14  \\
$UVM2$ & 2310 & 2.827 & 2.20  $\times 10^{-15}$ & 5.57   & 7.74  &  7.71  \\
$UVW2$ & 2120 & 3.167 & 5.71  $\times 10^{-15}$ & --     &  --   &  7.08  \\
\noalign{\smallskip}
\hline
\end{tabular}
 \begin{list}{}{}
 \item[$^{\mathrm{a}}$] The effective wavelength of the filter in \AA.
 \item[$^{\mathrm{b}}$] The ratio of $A_{\lambda_{e}}$, i.e. the extinction in
  magnitudes at the effective wavelength of the filter $\lambda_{e}$, to $A_{\rm V}$ (5500\AA)
  obtained from the interstellar reddening curve given by \citet{ccm} and updated by \citet{odonnell}.
 \item[$^{\mathrm{c}}$] The conversion factor from rate (Counts~s$^{-1}$) to flux
 (erg~cm$^{-2}$s$^{-1}$\AA$^{-1}$) from the OM in-orbit calibration (see SAS watch-out page).
 \item[$^{\mathrm{d}}$] The flux of PKS\,2005$-$489, corrected for galactic extinction, 
measured in the three epochs (Oct. 4, 2005; Sept. 26 and 28, 2005). The units are mJy.
 \end{list}
\end{table}

The data from the OM were processed with the 
tasks {\tt omichain} for the photometry and {\tt omfchain} for the timing analysis.
Since no variations are found in the light curves,
the whole exposure is used to derive the flux measurements.
The photometry was processed interactively with {\tt omsource} for every filter, 
to assure the use of point-source analysis procedures. 
A standard aperture of 6 pixels on the 2x2 binned images 
(12 for the unbinned images of the 2004 October data set) is used for all filters.
This corresponds to an aperture of 6$\arcsec$ for the optical filters (V, B, U).
The counts for the UV filters are extrapolated by the software, using the UV point-spread
function, to an aperture of 17$\arcsec$.5.   
These are the two apertures for which the OM count-rate-to-flux conversion is calibrated.
The OM images are affected by stray light from a bright star 
in the field of view, which increases the background near one side 
of the source region. Therefore background events are taken from both an 
annulus around the source and from two different circular regions 
at the same distance of the source from the main stray-light reflex.
Tests performed with different background sizes and locations
show that the photometry does not change by more than 1\%,
well below the systematic uncertainties of the flux conversion
(estimated at $\sim$10\%, see XMM-SOC-CAL-TN-0019).  

The source fluxes were obtained from the count rates using the 
OM conversion factors for white dwarfs\footnote{http://xmm2.esac.esa.int/sas/7.1.0/watchout/},
adding the 10\% systematic error in quadrature.
The fluxes were de-reddened for Galactic absorption 
using the extinction curve by \citet{ccm} with the updates by \citet{odonnell},  
and assuming $R_V[=A_V/E(B-V)]=3.1$. This is the average value for the Galactic diffuse ISM.
For the line of sight of PKS\,2005$-$489, a value of A$_{\rm B}=0.241$ is used 
\citep[from NED;][]{schlegel}, corresponding to A$_{\rm V(5500)}=0.182$.
The conversion factors, extinction ratios and the resulting source fluxes 
are reported for each of the filters in Table \ref{om}.

PKS\,2005$-$489 is hosted by a giant elliptical galaxy of 
total R magnitude 14.5 and half-light radius 5.7$\arcsec$
\cite[from HST snapshot observations,][]{scarpa}.
For the SEDs (Figs. \ref{xzoom} and \ref{seds}),
the OM fluxes were corrected for the contribution of the host galaxy.
The wavelength-dependent correction was determined using a template 
SED for elliptical galaxies \citep{silva98}, 
rescaled to the host-galaxy flux in the R band, and accounts 
for the given apertures. The OM fluxes are always dominated 
by the non-thermal emission, and the small contribution of 
the host-galaxy is only noticeable in the V and B filters. 

\begin{figure}
\centering
    \includegraphics[width=0.5\textwidth]{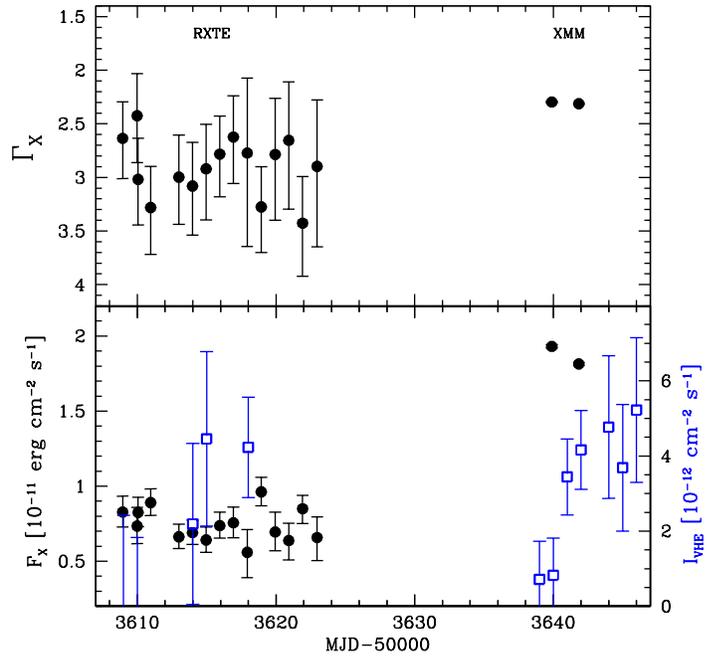}
   \caption{Simultaneous X-ray and VHE data taken by RXTE, \xmm~ and HESS
   in  August-October 2005. Upper panel: the X-ray photon index. The \xmm~ 
   values were obtained from fits above 2 keV, to be consistent with the RXTE energy range.
   Lower panel: the observed X-ray and VHE flux.  
   The X-ray data (filled circles) are energy fluxes integrated in the 2-10 keV band. 
   The HESS data (open squares) are integral photon fluxes above 400 GeV. 
   The vertical scales on the left and right are for the X-ray and HESS data, respectively.
   All errors are shown at the 1$\sigma$ confidence level.}
\label{xtev}
\end{figure}

\subsection{RXTE data analysis}
The {\it Rossi}XTE/PCA \citep{pca}  performed 15 snapshot observations of PKS\,2005$-$489 
between August 20 and September 19, 2005, yielding a total exposure of  23.4 ks.  
The observations were mostly done with PCU0 and PCU2.
The PCA STANDARD2 data were reduced and analyzed 
with the standard FTOOLS routines in HEASOFT V6.3.2, 
using the filtering criteria recommended by the RXTE Guest Observer Facility.
The new SAA history file and parameters for the background calculation provided 
in September 2007 were used,
which correct a bug in the calculation of the faint background 
model\footnote{Details at http://www.universe.nasa.gov/xrays/programs/rxte/pca/
doc/bkg/bkg-2007-saa/}.
Only the top-layer events were processed and only the PCU2 data were considered, 
to ensure a more accurate spectral measurement.
The average net count rate in the 3-15 keV band is $0.54\pm0.02$ cts\,s$^{-1}$\,pcu$^{-1}$.
The RXTE spectra were extracted and fitted separately for each pointing, 
and summed together to obtain the average spectrum.
Each spectrum is well-fit by a single power-law model.

\section{XMM-Newton/RXTE results}
\subsection{Flux variability}
The 2004 and both 2005 EPIC light curves
contain no evidence of flux variability within each exposure, and 
have average net count rates in the MOS2 camera of
$1.14\pm0.04$, $6.38\pm0.08$, and $6.15\pm0.08$ counts s$^{-1}$, respectively.
The probability of constant emission is P$_{\rm const}\geq 0.99$, 0.92 and 0.99 
for the three epochs, respectively.  Similarly, no 
flux variability is observed in the OM data for any of the observed filters.
No significant flux variations (P$_{\rm const}\geq0.76$) are detected 
during the RXTE observations (see Fig. \ref{xtev}) as well. However, 
night-by-night flux variations 
of a few tens of percent
cannot be excluded, due to low statistics in the RXTE measurements.

Although no variability is found within any of the \xmm~ exposures
or within the complete RXTE sample, strong flux variations
are found on longer time scales. The 2-10 keV flux varies by a factor of $\sim$16 between
2004 and 2005.  In the UV and optical bands the flux increases by 
$\sim$40\% and $\sim$20\%, respectively, between 2004 and 2005.
A smaller (factor of 2.5), but significant, change in the 2-10 keV flux
also occurs during the 17 days between the final RXTE and first \xmm~ 
pointing in September 2005.

\subsection{X-ray spectrum}
Since there is no variability in the X-ray flux or hardness ratio,
during each \xmm~ exposure, a single X-ray spectrum was extracted 
for each of the three pointings  (see Table \ref{xfits} and Fig. \ref{xplots}).
The spectra from the individual RXTE exposures 
do not vary significantly (see Fig.~\ref{xtev})
and are consistent with the fit to the complete RXTE sample reported
in Table \ref{xfits}.
In each case (\xmm~ and RXTE), the Galactic absorption is fixed to
$N_{\rm H}=3.93\times10^{20}$ cm$^{-2}$, as recently determined 
from the LAB survey \citep{kalberla}. It is important to note that
there is a discrepancy between the LAB survey estimate 
and the value obtained from the HI maps by \citet{dl90} 
(DL, $N_{\rm H}=5.08 \times10^{20}$ cm$^{-2}$).  
The LAB survey value is adopted because
it provides better residuals, is consistent with the 
results when $N_{\rm H}$ is left as free parameter ($(3.96\pm0.17) \times10^{20}$ cm$^{-2}$),  
and is close to the value previously adopted for the \sax results on the large flare of 1998
\citep[$4.2\times10^{20}$ cm$^{-2}$,][]{gt}.
Using the DL value significantly changes the 
slope of the source spectrum below 1 keV.
Therefore both cases are considered for  
interpreting the fits results, as additional systematic uncertainty.

In all three data sets, the EPIC spectra are better fit by a broken power-law model
with respect to a single power law, with high significance (F-test $>99.9$\%).
Remarkably, the parameters reveal an inverted broken power-law spectrum,
where the slope in the soft X-ray band is steeper than in the hard X-ray band
($\Gamma_2<\Gamma_1$).
\begin{figure}
\centering
    \includegraphics[width=0.5\textwidth]{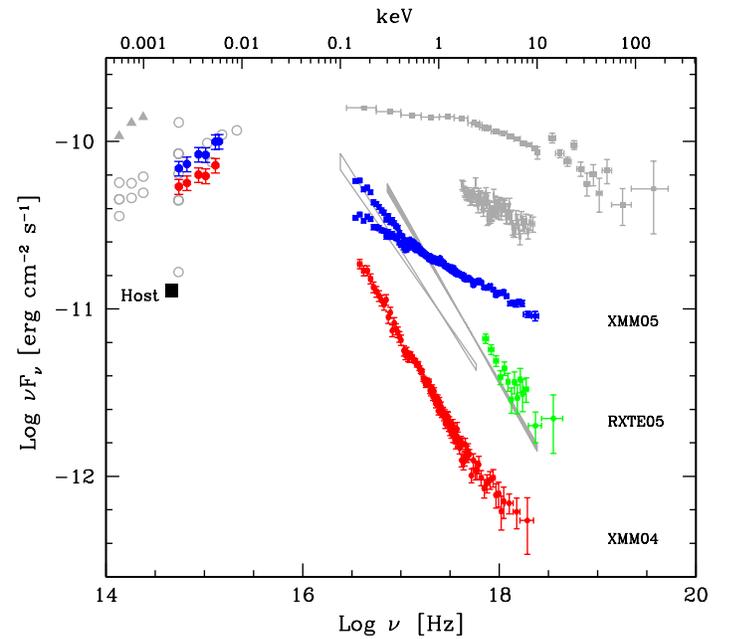}
   \caption{SED of PKS\,2005$-$489 in the frequency range 
   covered by the \xmm~ and RXTE observations (filled circles).  Archival X-ray spectra
   are shown along with those 
   measured by \xmm~ on Oct. 4, 2004 (red points), RXTE in Aug-Sept 2005 (green points) 
   and \xmm~ on Sept. 26, 2005 (blue points). For the latter, 
   two  fits are plotted, showing the effects of using two different
   estimates of the Galactic column density 
   ($N_{\rm H}=5.08$ or $3.93 \times10^{20}$ cm$^{-2}$; see text).
   For better visibility, the \xmm~ spectrum of Sept. 28, 2005 is not plotted
   as it would appear  superposed on the Sept 26 data.  
   The flux of the host galaxy in the R band is plotted as black square. 
   Historical data are shown in gray \citep[from NED and][]{gt}. 
   In the X-ray range, they correspond to (from top to bottom):
   the \sax data of the 1998 flare \citep{gt}, the \sax 1996 data \citep{padovani},
   the \swift spectrum in April 2005 \citep[filled bow-tie,][]{massaro08}, 
   and the 1992 \rosat spectrum \citep[open bow-tie,][]{comastri}.   
   } 
\label{xzoom}
\end{figure}

In the 2004 dataset, the X-ray spectrum is characterized by the lowest flux 
ever recorded from this source (see Fig. \ref{xzoom})
and a very steep slope of $\Gamma\simeq3$ up to $\sim$3 keV.
At higher energies, there is evidence of a moderate hardening trend.
The trend is consistent between the MOS and PN spectra,
but is only significant in the PN data thanks to  better 
statistics.
This hardening can also be reproduced by the sum of two 
power-law models, the second having a harder index.
The slope of this second power-law function is not well constrained.
However, assuming a fixed value of $\Gamma=1.5$, 
as could be expected for the inverse Compton emission of low-energy electrons,
a good fit ($\chi^2_r=1.07$ for 283 NDF) is obtained with 
a flux normalization at 1 keV for the second power-law function 
at $2.8\pm0.7$\% of the first power law. 
In this case, the two power-law functions would cross at about 20 keV. 
Using the high  
$N_{\rm H}$ has minimal impact on this result, 
steepening the low-energy slope by only $+0.1$.  

The difference between the EPIC spectra 
of September 26 and 28, 2005  is negligible.  However, the spectrum 
from either of the 2005 observations is significantly
harder and brighter than in 2004 (Fig.~\ref{xzoom}).
Either 2005 spectrum is represented well by a pure power-law model with $\Gamma=2.3$
from $\sim$1 to 10 keV.  
At lower energies, each has a slightly steeper slope.
Although modest ($\Delta\Gamma\lesssim0.1$), the spectral break 
is significant (F-test $>99.99$\%) 
and is not caused by the extension of the MOS data at low energies, 
or by the cross-calibration between the detectors. The broken power-law model
is also statistically required for the PN spectrum alone (F-test $>99.99$\%), 
as well as for the MOS spectra fitted separately.
It should be noted however that the precise location of the break
is affected by systematic more than statistical errors.
In Table 3,  the break energy for the fit with 
the overall minimum $\chi^2$ is reported,  but there are other similar 
local minima between 0.7 and 2.2 keV, 
whose hierarchy in $\chi^2$ can change according to the calibration
of the effective area near the instrumental Si and Au edges (at 1.7 and 2.1 keV).
The true break therefore should be considered more realistically inside the range 0.7--2.2 keV.
This does not affect the values of the soft and hard slopes in a relevant way
(they do not change by more than $0.04$ and $0.02$, respectively).

The amount of the spectral break, however, depends on the adopted value of
the Galactic column density.  Using the DL value instead
of the LAB survey value, the soft X-ray slope becomes steeper ($\Gamma_1=2.6$),
making the break more pronounced ($\Delta\Gamma\simeq0.3$).  
The two cases are shown in Fig. \ref{xzoom}, 
where the spectra are plotted together with the other data sets
and archival data. Allowing instead the column density to be lower than all
available estimates, a single power-law spectrum is obtained with 
$N_{\rm H}=3.76\pm0.08$ cm$^{-2}$ ($\chi^2_r=1.226$ for $859$ NDF), though a broken power-law
model still seems to provide a better representation (F-test $>95$\%).
\begin{table}[t]
 \caption{Best power-law fits to the HESS spectra corrected for 
 EBL absorption following \citet{franceschini}.}  
 \label{deabs}
     \centering
      \begin{tabular}{l c c c}
	 \hline\hline
	 \noalign{\smallskip}
	  Epoch  & $\Gamma_{\rm int}^{\mathrm{a}}$ & $\nu F_{\nu} ({\rm 1 TeV})^{\mathrm{b}}$ & $\chi^2$\,(NDF)  \\
     		 &	    & [$10^{-12}$ erg cm$^{-2}$\,s$^{-1}$] &   \\
	 \noalign{\smallskip}
	 \hline
	 \noalign{\smallskip}
     	 2004 &  $3.1\pm0.4$ & $1.8\pm0.4$ & 6.4\,(7)  \\
     	 2005 &  $2.6\pm0.2$ & $2.7\pm0.4$ & 13.2\,(6)  \\
     	 2006 &  $2.4\pm0.2$ & $2.9\pm0.4$ & 4.3\,(7)  \\
     	 2007 &  $3.1\pm0.7$ & $1.2\pm0.9$  & 0.5\,(2)  \\
	 \noalign{\smallskip}
	 \hline
	 \noalign{\smallskip}
     	 2005 7-8$^{\mathrm{c}}$ & $3.2\pm0.4$ & $2.4\pm0.6$ & 9.2\,(4) \\
     	 2005 \,\,9$^{\mathrm{c}}$    & $2.5\pm0.3$ & $2.9\pm0.5$ & 4.7\,(6) \\
	 \noalign{\smallskip}
	 \hline
	 \noalign{\smallskip}
     	 Total &  $2.69\pm0.16$ & $2.34\pm0.20$ & 8.9\,(8) \\
	 \hline
    \end{tabular}
 \scriptsize
\vspace{-0.2cm} 
 \begin{list}{}{}
  \item[$^{\mathrm{a}}$] Only the statistical errors are shown.
  \item[$^{\mathrm{b}}$] The normalization at 1 TeV, in $\nu F_{\nu}$ units.
  \item[$^{\mathrm{c}}$] The HESS spectrum in 2005, extracted in two different epochs 
  (see Fig. \ref{nightly_plots}):  before the XMM pointings (observing periods 
  2005, 7-8) 
  and during the XMM epoch (period 2005, 9).  
 \end{list}
\end{table}

The 2005 RXTE observations yield a very steep ($\Gamma=2.9$) spectrum from 3-15 keV.
This spectrum is almost parallel to the 2004 \xmm~ spectrum but with 
$\sim$6 times higher flux. The RXTE photon index and flux are 
in good agreement with the values measured by \swift in April 2005 \citep{massaro08}.
This suggests that the hardening of the X-ray spectrum of PKS\,2005$-$489
occurred in September 2005, between the RXTE and \xmm~ pointings.

\section{Discussion}
\begin{figure*}[t]
\centering
    \includegraphics[width=0.95\textwidth]{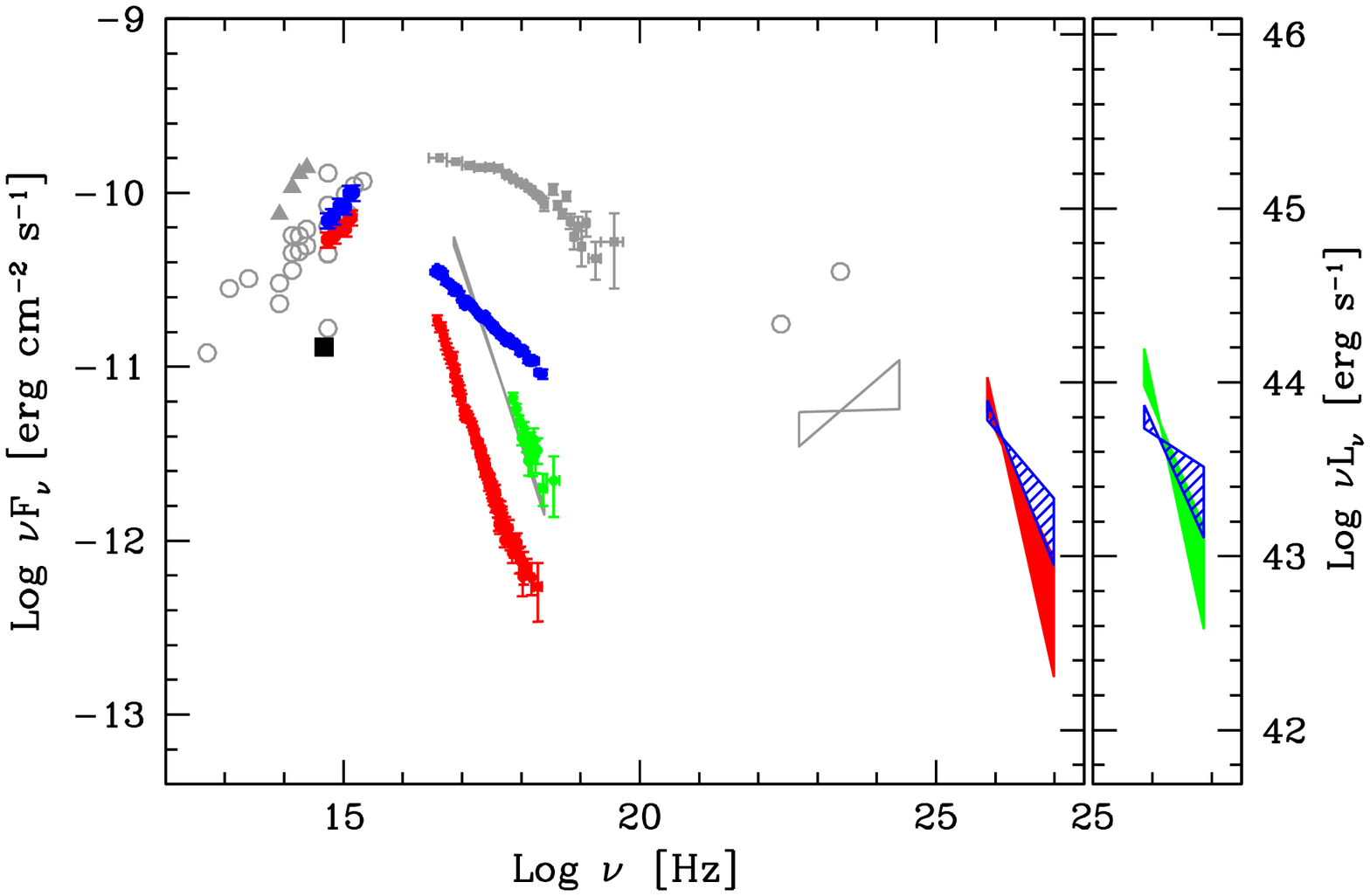}
   \caption{SED of PKS\,2005$-$489, in different epochs and states. 
   The \xmm~ and RXTE data are plotted as in Fig. \ref{xzoom}, 
   together with historical data \citep[shown in grey; from][]{gt,massaro08}.
   The recent {\it Fermi}-LAT spectrum (0.2$<$E$<$10 GeV) from Aug--Oct 2008 is also plotted, 
   for reference \citep[open bow-tie;][]{lbas}. 
   For clarity, the HESS spectra (E$>$300 GeV) are plotted as bow-ties, 
   and have been corrected for EBL absorption with the model by \citet[][see Table 5]{franceschini}. 
   Main panel: the HESS time-averaged spectrum from 2004 (red filled bow-tie) 
   and 2005 (blue hatched bow-tie). Right mini-panel: the HESS spectrum in 2005, divided in two subsets:
   before the XMM pointings (periods 07-08/2005 in Fig. \ref{nightly_plots}; green filled bow-tie) 
   and during the XMM epoch (period 09/2005 in Fig. \ref{nightly_plots}; blue hatched bow-tie).
   Same colors 
   associate optical, X-ray and VHE datasets corresponding to same epochs.} 
\label{seds}
\end{figure*}

The spectral energy distribution (SED) of PKS\,2005$-$489,
assembled with the data taken in 2004 and 2005, is shown in Fig. \ref{seds}.
The VHE $\gamma$-ray spectra are corrected for the energy-dependent attenuation
caused by interactions with the optical-IR photons of the diffuse extragalactic 
background light (EBL).
The correction was performed adopting the EBL model of \citet{franceschini}.
This model takes into account the most recent results on galaxy emission and evolution,
and agrees with both the upper limits on the EBL derived from blazar spectra
\citep{nature_ebl,0229} and the lower limits given by galaxy counts, 
at both optical and infrared wavelengths \citep{madau,fazio}.
The fits to the absorption-corrected spectra are reported in Table \ref{deabs},
and plotted in Fig. \ref{seds}. 

The simultaneous observations do not show evidence 
of strong changes in the location
of the SED peaks over the years, with respect to the historical values.
The hard optical-UV spectrum from the OM photometry
and the steep X-ray spectrum above 0.1 keV
locate the synchrotron peak between the two bands, at approximately
0.5$\times10^{16}$ Hz.
At VHE, the steep spectra constrain the peak of the $\gamma$-ray emission 
 to energies $<$0.2 TeV.
The low flux in the VHE band does not enable a determination of the spectral properties 
in exactly the same (short) observing windows of the X-ray observations.
However, given the lack of significant VHE variability in 2004 or 2005,
the monthly or yearly average spectra can be considered a reasonable approximation 
to the VHE spectra during the epoch of the X-ray observations. 

In 2004,  PKS\,2005$-$489 was in a very low state in both the X-ray and VHE bands.
The spectral slopes are also similar ($\Gamma\sim3$) for the X-ray and VHE emission.
The X-ray flux around few keV is more than two orders of magnitude less than
the \sax flux during the active state of 1998.  The slight spectral hardening 
towards higher X-ray energies is consistent with the onset of the concavity 
expected from the emergence of the inverse-Compton emission in the SED. 
This emission, produced by low-energy electrons, is usually not visible
in the X-ray spectra of HBL because of the dominance of
the synchrotron emission from high-energy electrons. In this data set, 
its presence becomes detectable because of the steepness and very low flux 
of the synchrotron (X-ray) spectrum. A similar behavior has recently been observed 
in the HBL PKS\,2155-304 during a low state \citep{foschini_rem,zhang08}.  
This transition between steep synchrotron and flat
inverse-Compton spectra  
is instead fully visible 
inside the observed X-ray passband for intermediate BL Lac objects 
\citep[e.g. ON\,231, see][]{on231}.

In 2005, the X-ray spectrum changes radically.
\swift and RXTE measurements made in April and August-September 2005, respectively,  
show a large increase in flux  but little change in the spectral slope
compared to 2004.  Shortly after the RXTE observations, a
much harder spectrum ($\Delta\Gamma_{\rm X}\simeq0.7$)
and a brighter state than in the previous epochs is found during \xmm~ observations.
Most interestingly, the data require a spectral break with a steeper spectrum 
towards lower energies. This break is more apparent 
when a higher galactic N$_{\rm H}$ is used 
(e.g. the DL versus LAB survey values; see Fig. \ref{xzoom}),
but even when considering a single power-law shape with an ad-hoc N$_{\rm H}$, the 
extrapolation of the X-ray spectrum down to lower frequencies does 
not match the UV-optical data, which are still rising with frequency.
A hump with a steeper soft X-ray slope is needed to smoothly 
connect with the optical-UV spectrum, and this feature
indicates the presence of two particle populations, since it
cannot be easily reproduced by synchrotron radiation with a single electron distribution.

At VHE, the comparison between the 2004 and 2005 spectra indicates 
a hardening trend as well, 
though with  marginal significance ($\sim$1 sigma; see Table \ref{deabs})
due to the large errors on the HESS measurement.
However, if this trend is real, and if the VHE spectrum is indeed produced by the same 
particles as are responsible for the X-ray emission 
(as suggested by the similar spectral slopes
in 2004-2005 and by the location of the synchrotron 
peak\footnote{If the VHE electrons were emitting by synchrotron at energies 
well below the X-ray range, they would correspond to the synchrotron peak, around $10^{16}$ Hz.     
Since the SSC scattering at the peak would be in the Thomson regime for this frequency, 
the gamma-ray peak should then be inside the VHE band, which is not observed.}),
a similar behavior is expected. Namely, the VHE spectral hardening
should have occurred after the RXTE observations, during the epoch of the \xmm~ pointings,
while the 2005 July-August VHE spectrum should be similar to the 2004 spectrum (modulo normalization). 

To test this hypothesis, the 2005 HESS data set was divided into two subsets.
The first consists of all VHE data before the \xmm~ pointings 
(2005 July-August dark periods),
and the second contains all 2005 HESS data taken during or after the pointings 
(2005 September dark period).
Table~\ref{deabs} shows the results of power-law fits to the spectra from these subsets, 
after correction for the absorption of VHE photons on the EBL.
The HESS spectrum in the XMM epoch is indeed harder ($\Delta\Gamma_{\rm int}=0.72\pm0.47$) 
than the pre-XMM spectrum,  and the statistical significance of the spectral 
change increases ($\sim$2$\sigma$) with respect to the comparison between yearly spectra.
In addition, the pre-XMM spectrum  has the same slope as the 2004 HESS spectrum.
The HESS data therefore  are fully consistent with a behavior in the VHE band 
that mirrors the spectral variations seen in the X-ray band, 
as expected if both emissions are produced by the same electrons.
Interestingly, the harder VHE state continues in 2006. 
Because of the limited
exposure in 2007, the VHE statistics are too  poor 
to draw any meaningful conclusions. 

The most remarkable feature of the VHE emission, however, is the lack of 
flux variability. Comparing the average states in 2004 and 2005,
the integrated energy flux (e.g. for reference in the decade 0.3-3 TeV)
does not vary by more than 40\% between  2004 and 2005, 
and is consistent with a constant flux. In contrast, the X-ray flux 
increased by a factor of $\sim$16 in the 2-10 keV band (see Table \ref{xfits}). 
The optical-UV emission, which is close to the synchrotron peak
and which typically provides the target photons for the IC scattering 
in the VHE band, shows an increase as well ($\sim$40\%).
In a homogeneous SSC scenario, if the VHE emission is produced 
by the same electrons emitting in the X-ray band 
(as indicated in this case by the spectral behavior),
a fresh injection of electrons should make the VHE flux increase
at least linearly with the X-ray flux. 
The reason is that the energy density of all possible seed photons
for the IC scattering has increased or is constant between the two epochs.
For the VHE flux to remain constant, a corresponding decrease in the seed photons 
energy density is required.

The X-ray and VHE data taken together, therefore, suggest 
that a new jet component is emerging in the SED of PKS\,2005-489, 
which is physically separated from the main emitting region.
The emission at the synchrotron peak of this new component should be lower, 
remaining hidden below the observed SED, while its harder spectrum at high energies
emerges in the hard-X band.
The electrons of this new component would not see the 
energy density of the observed synchrotron peak,
but the lower energy density of their self-produced
synchrotron peak instead.

\section{Conclusions}
PKS\,2005$-$489 is detected at VHE in each of the four
years it was being observed by HESS (2004--2007).  The 2005--2007
data clearly confirm the VHE discovery reported by
HESS \citep{hess_discovery} and quadruple the statistics
of the initial spectrum measurement.   Re-analysis of the previously
published 2004 HESS data, using the improved calibration of the detector's energy
scale, results in a $\sim$3 times higher flux and a similar photon index.

The measured VHE $\gamma$-ray flux is low ($\sim$3\% Crab)
and only shows weak variations on time scales ranging from days to years.
The observed time-averaged (2004--2007) VHE spectrum is 
soft, with a photon index $\Gamma = 3.20\pm0.16$.
Although evidence of VHE spectral variations is marginal by itself, 
the VHE spectrum seems to track the X-ray slope variations when 
multi-wavelength coverage is available.
Observations performed with \xmm~ and RXTE in 2004 and 2005 
reveal remarkable changes in the X-ray spectrum, 
but without shifting the location of the synchrotron 
peak with respect to historical observations.

Interpreting these
measurements along with the HESS data suggests the emergence of a 
new jet component in the SED that is characterized by a harder 
electron spectrum.  This component must be separate:
its particles cannot interact with
the synchrotron photons of the observed SED peak, otherwise
higher VHE fluxes than observed would be implied.

PKS\,2005$-$489 is found overall in a very low state, 
in both the X-ray and VHE bands, during the observations
presented in this article. PKS\,2005$-$489 has historically 
demonstrated a 100$\times$ dynamical range in the X-ray band.
Thus, dramatically higher VHE fluxes ($10^2-10^4\times$) 
can be expected in the future, unless such an increase 
in the X-ray flux is counter-balanced by a strong ($>$10$\times$) 
and simultaneous increase in the blazar's magnetic field.
Further monitoring of this object is highly encouraged,
as it is one of the few HBL easily detected at VHE during 
low states and has the potential for extreme brightness 
and variability. 

The results presented here confirm the strong diagnostic potential
of coordinated optical--X-ray--VHE observations. Future studies
can be significantly improved by incorporating data from the recently launched 
Fermi $\gamma$-ray satellite.  The Fermi data will
provide information on the lower energy side of the inverse-Compton peak,
enabling contemporaneous measurements of both sides of each blazar hump.

\begin{acknowledgements}
The support of the Namibian authorities and of the University of Namibia
in facilitating the construction and operation of HESS is gratefully
acknowledged, as is the support by the German Ministry for Education and
Research (BMBF), the Max Planck Society, the French Ministry for Research,
the CNRS-IN2P3 and the Astroparticle Interdisciplinary Programme of the
CNRS, the U.K. Science and Technology Facilities Council (STFC),
the IPNP of the Charles University, the Polish Ministry of Science and 
Higher Education, the South African Department of
Science and Technology and National Research Foundation, and by the
University of Namibia. We appreciate the excellent work of the technical
support staff in Berlin, Durham, Hamburg, Heidelberg, Palaiseau, Paris,
Saclay, and in Namibia in the construction and operation of the
equipment. The article is based on observations obtained with XMM-Newton, 
an ESA science mission with instruments and contributions directly 
funded by ESA Member States and NASA.
The authors thank the RXTE team for support during the ToO trigger.
This research has made use of the NASA/IPAC Extragalactic Database (NED) 
which is operated by the Jet Propulsion Laboratory.
\end{acknowledgements}

\bibliographystyle{aa} 
\bibliography{pks2005_mwl} 
\end{document}